\documentclass[12pt,a4paper]{article}
\usepackage{natbib}

\usepackage{graphicx}
\usepackage{bm}
\usepackage{amsmath}
\usepackage{subfigure}
\usepackage{multirow}
\usepackage{tabularx}
\usepackage{moreverb,url}
\usepackage{subfigure}
\usepackage{setspace}
\begin{document}

\title{\vspace*{-1.5cm} A multiple inflated negative binomial hurdle regression model: analysis of the Italians' tourism behaviour during the Great Recession \vspace*{5mm}}

\author{Chiara Bocci$^*$    \and
        Laura Grassini$^*$  \and
        Emilia Rocco\thanks{Department of Statistics, Computer Science, Applications ``G. Parenti'', University of Florence, Italy. 
             e-mail:  {\tt chiara.bocci@unifi.it, laura.grassini@unifi.it, emilia.rocco@unifi.it} }
}



\date{}

\maketitle

\begin{abstract}
We analyse tourism behaviour of Italian residents in the period covering the 2008 Great Recession.
Using the \emph{Trips of Italian Residents in Italy and Abroad} quarterly survey, carried out by the Italian National Institute of Statistics, we investigate whether and how the economic recession has affected the total number of overnight stays.
The response variable is the result of a two-stage decision process: first we choose to take a holiday, then for how long. Moreover, since the number of overnight stays is typically concentrated on specific lengths (week-end, week, fortnight) we observe multiple peculiar spikes in its distribution.
To take into account these two distinctive characteristics, we generalise the usual hurdle regression model by specifying a multiple inflated truncated negative binomial distribution for the positive responses.
Results show that the economic recession impacted negatively on both components of the decision process and that, by controlling for the inflated nature of the response variable's distribution, the proposed formulation provides a better representation of the Italians' tourism behaviour in comparison with non-inflated hurdle models. 
Given this, we believe that our model can be a useful tool for policy makers who are trying to forecast the effects of new targeted policies to support tourism economy.
\end{abstract}

\section{Introduction}
\label{sec:1}

During the years 2008-2013, consumption expenditures of Italian households was harshly hit by the Great Recession \citep{ISTAT2014} with a remarkable reduction of purchasing power (-10.4\% between 2007 and 2013).
In that period, Italian households showed a reduction in tourism expenditure and a change in travel behaviour as well.
The expenditure share devoted to accommodation facilities passed from 2.8\% in 2010 to 2.3\% in 2013 and the annual decrease in the number of trips by resident was nearly -12\% in 2010, -19\% in 2013.
Only in 2015, for the first time after seven years, there has been a remarkable increase (+13.5\%).

Objective of our study is tourism behaviour of the Italian residents and, in particular, we analyse Italians' participation in tourism in the period covering the recent economic recession.
Using data from the survey on \emph{Trips of Italian Residents in Italy and Abroad}, carried out quarterly by the Italian National Institute of Statistics (ISTAT), we investigate whether the propensity in tourism participation  (i.e., the probability of having at least one holiday trip with at least an overnight stay in a quarter) and the intensity of participation  (i.e., the sum of the length of stay of all holiday trips taken in a quarter) have changed over the period of analysis. 

The theoretical framework used for the joint analysis of these two aspects is the hurdle model, a modified count data model which allows to consider the response as result of a two-stage decision process: at first a person decides whether to take a holiday and then, conditionally to a positive decision, he decides the length of the holiday.
In a general hurdle model, a binary model is used to represent the binary outcome of whether a count variable has a zero or a positive realisation and then the positive realisations are modelled by a truncated-at-zero count data model.
Various specifications can be adopted for the truncated-at-zero model depending on the distribution of the positive realisations.
Given their flexibility, the hurdle models have been widely used in several contexts of health and economic studies, and a few applications of this method can be found in tourism analysis as well \citep{Hellstrom2006, Bernini2015, Bernini2016, BotoGarcia2019}.

A noteworthy concern in the analysis of the number of overnight stays is the presence of multiple spikes in its distribution.
That occurrence is due to the propensity to take a holiday in typical day blocks (e.g. week-end, one week, two weeks, etc.), which in turn produces a concentration of the total number of overnight stays on certain values, known as inflated values.
Some authors have treated this problem by re-defining the response variable into two or more classes and then applying a logit or a multinomial model \citep{Alegre2006, Nicolau2004}; others adopted a latent class approach \citep{Alegre2011} or employed a quantile regression model \citep{Salmasi2012}.

We take a novel approach, not yet adopted in the context of tourism analysis: the truncated-at-zero model for the positive responses is specified as a multiple inflated truncated negative binomial model, that is a finite mixture of a zero-truncated negative binomial and a set of degenerate distributions on the inflated values, with the mixture probabilities modelled through a multinomial logit model.
Even considering a wider literature, at the best of our knowledge the actual specification of a hurdle model with a multiple inflated distribution for the positive responses, even if theoretically feasible, has not been presented before.

Results show that the economic recession impacted negatively on both components of the decision process and that, by controlling for the inflated nature of the response variable distribution, the proposed formulation provides a better representation of the Italians' tourism behaviour in comparison with non-inflated hurdle models. 
In particular, by using a multiple inflated hurdle model we are able not only to identify the determinants of the phenomenon under study, but also to correctly fit the distribution of the total number of overnight stays, even in presence of extremely inflated values which are usually under-predicted by standard models.
Given this characteristic, we believe that the use of multiple inflated hurdle models can produce results which could be useful for policy makers in evaluate how the Italians would react to the implementation of targeted tourism policies.

The paper is organized as follows.
Section \ref{sec:3} describes the database and discusses the characteristics of the response variable.
Section \ref{sec:4} presents the theoretical model used for the analysis and discusses its properties and usefulness for the aim of the study.
Results of the empirical analysis are presented in Section \ref{sec:5}, and the last section concludes with the discussion of the main findings.

\section{Data}
\label{sec:3}

The analysis employs a pooled time series cross-section database of Italian residents in the period 2004-2013, which covers the last economic recession that has seriously affected Italian households.
Data comes from the household survey on \emph{Trips and Holidays of Italian Residents in Italy and Abroad}, which is the main statistical source of demand-side tourism data available in Italy.
It is currently carried out by ISTAT for responding at the EU Reg.692/2011, and it collects information about domestic and outbound travels of the Italian residents.

From 1997 to 2013, it has been conducted quarterly on a national annual sample of about 14,000 households (about 3,500 per quarter), comprising an annual total of about 32,000 individuals.
Each year, data are collected for the periods January-March, April-June, July-September and October-December.
In each quarter and for each individual, information on travels with at least one overnight stay concluded during the quarter, made for any main purpose, are recorded.
Tourism trips are classified into business and holiday trips.

In addition, socio-demographic characteristics of all household components are recorded: age, gender, region of residence, education level, marital status, occupational status and professional position.
It should be noted that this information is collected for all individuals, regardless of their being traveller or not.
Therefore the survey data allows to identify the share and characteristics of both tourism participants and non-participants.
Unfortunately, these characteristics do not include any information about the individuals' economic status.

For the participants the survey offers also an in-depth insight about their tourism behaviour in terms of number of trips, nights spent and characteristics of the trip, but provides no information about tourism expenditure (although surveyed for the Tourism Satellite Account it is not provided for research purpose).

From 2014, the \emph{Trips and Holidays} survey has become a focus included in the \emph{Households Budget Survey} and deep changes have been introduced in every stage of the survey process.
Therefore, since the two sources cannot be appropriately linked together, we stop our analysis at 2013.
Moreover, given the adoption of the Euro currency occurred in 2002, we start the analysis from 2004.

\begin{table} [!htb]
\caption{Descriptive statistics of response variable} \label{tab_descrittive}
\begin{center}
  \resizebox{0.75\linewidth}{!}{
   \begin{tabular}{lrr}
    \hline\noalign{\smallskip}
      & Number of overnight & Positive number of overnight\\
      & stays in a quarter & stays in a quarter \\
    \noalign{\smallskip}\hline\noalign{\smallskip}
    Min         & 0 & 1   \\
    Q1          & 0 & 3 \\
    Median      & 0 & 6    \\
    Q3          & 0 & 12 \\
    Max         & 270  & 270 \\
    Mean        & 2.01 & 9.59   \\
    Variance    & 43.29 & 134.19  \\
    N           & 313368  & 65569   \\
   \noalign{\smallskip}\hline\noalign{\smallskip}
  \end{tabular}}\\
\end{center}
\end{table}

As we are interested in studying the factors that may influence individual tourism behaviour, our unit of study is the individual.
We limit the analysis to holiday trips and, since children's tourism choices are not individually made, we consider only persons at least 15 years old.

We define our study variable as the total number of overnight stays in a quarter, obtained by summing the length of stay of each holiday trip made by an individual in that quarter (the variable is set at zero for an individual who has not travelled).
The variable's descriptive statistics presented in Table \ref{tab_descrittive} show a prevalence of zero values: almost 80\% of the sampled units are tourism non-participants.
If we expand the data to the whole Italian population by using the expansion factor provided by ISTAT, we can estimate that only about the 24\% of Italians (at least 15 years old) made on average at least a holiday trip in a quarter.
Even in the summer quarter (July-September), this percentage reaches only 42\%.

\begin{figure}[!tb]
\begin{center}
 \includegraphics[width=.9\textwidth]{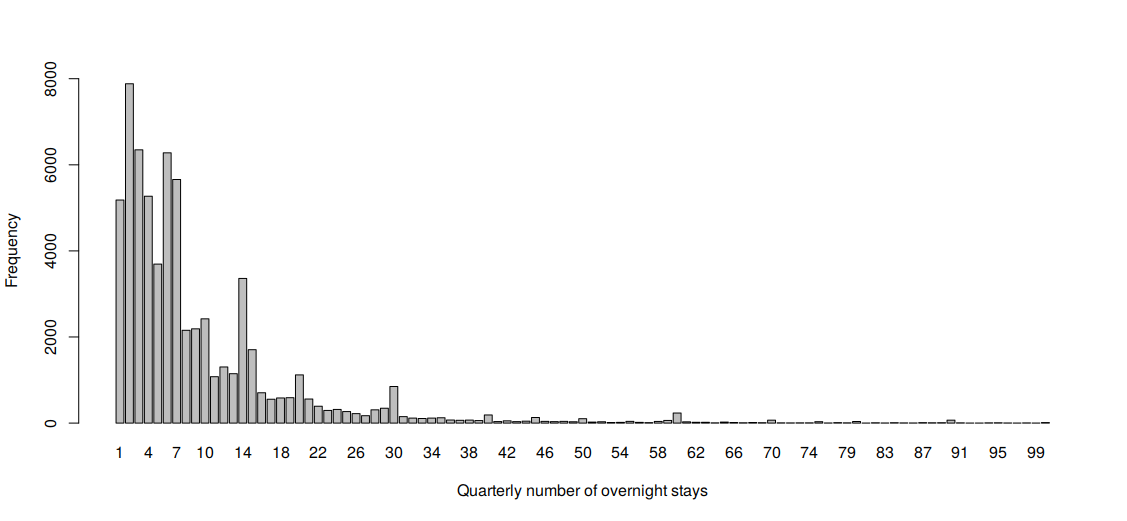}
\end{center}
\caption{Histogram of the positive number of overnight stays in a quarter} \label{istogramma}
\end{figure}

Considering the positive number of overnight stays, from Table \ref{tab_descrittive} we can see that the variable is highly skewed and overdispersed (the variance is almost 14 times the mean).
This is confirmed by Figure \ref{istogramma}: when the number of overnight stays increases its frequency quickly decreases, but the distribution has a long tail of low-occurrence values.
From Figure \ref{istogramma} we can observe multiple spikes in its distribution: the observed positive values of overnight stays are concentrated on specific values, like 2, 6, 7, 14, 15, 20 and 30 nights. 

\begin{figure}[!htb]
\begin{center}
 \includegraphics[width=\textwidth]{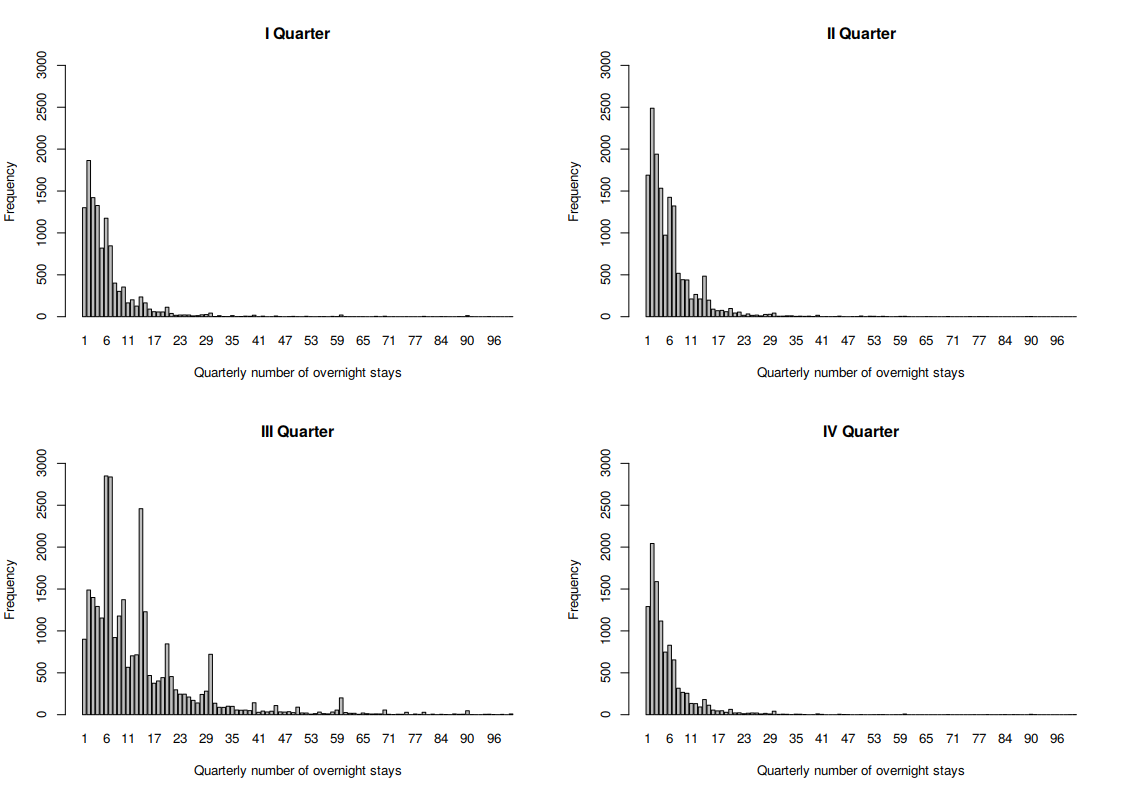}
\end{center}
\caption{Histogram of the positive number of overnight stays by quarter} \label{istogramma2}
\end{figure}

Finally, as expected, seasonality plays an important role in characterising tourism behaviour.
The proportion of Italians who made on average at least a holiday trip in a quarter and the average number of overnight stays are significantly higher in the third quarter than in the other quarters.
More important, as we can see in Table \ref{tab_descrittive_trim} and Figure \ref{istogramma2}, the distribution of the positive number of overnight stays is completely different in the third quarter from that of other quarters: it is more variable, with a longer tail and with a larger number of inflated values. In addition, there is a higher concentration on the inflated values.

\begin{table} [!tb]
\caption{Descriptive statistics of the positive number of overnight stays by quarter} \label{tab_descrittive_trim}
\begin{center}
  \resizebox{0.9\linewidth}{!}{
   \begin{tabular}{lrrrr}
    \hline\noalign{\smallskip}
    Positive number of & I Quarter & II Quarter & III Quarter & IV Quarter \\
    overnight stays & \emph{Jan. - Mar.} & \emph{Apr. - June} & \emph{July - Sept.} & \emph{Oct. - Dec.} \\
    \noalign{\smallskip}\hline\noalign{\smallskip}
    Min         & 1     & 1     & 1     & 1   \\
    Q1          & 2     & 2     & 6     & 2   \\
    Median      & 4     & 4     & 10    & 4   \\
    Q3          & 7     & 7     & 17    & 7   \\
    Max         & 180   & 270   & 250   & 240  \\
    Mean        & 6.62  & 6.39  & 13.80 & 5.80 \\
    Variance    & 77.16 & 67.85 & 182.65 & 70.43 \\
    N           & 11492 & 14983 & 28741 & 10293 \\
   \noalign{\smallskip}\hline\noalign{\smallskip}
  \end{tabular}}\\
\end{center}
\end{table}

\section{Methodology}
\label{sec:4}

Since the response variable is discrete and non-negative, we refer to count data models. The most common models in literature are the Poisson and the negative binomial regression models.

One of the basic assumptions of these models is that both zero and positive values of the response variable come from the same data generating process.
However, as it frequently occurs when analysing socio-economic phenomena, our data do not adhere to this assumption.
In fact, it makes sense to assume that a person firstly decides whether or not to take a holiday (i.e. whether or not to participate in tourism), and then, conditionally to a positive decision, he decides the number of overnight stays.
In such a situation it seems opportune to firstly separate participants from non-participants, zeroes from non-zeroes, through a binary model and then to model the positive responses using a truncated-at-zero count data model.

This assumption is typical of the hurdle model, in which the two processes generating zeros and positive values are not constrained to be the same \cite{Cameron2013}.
Firstly, a binomial probability governs the binary outcome of whether a count variate has a zero or a positive realisation and then, if the hurdle is crossed (i.e. the realisation is positive), the conditional distribution of the positives is governed by a truncated-at-zero count data model.
Such a conditional setting enables the interpretation of covariate effects through event incidence and frequency in the respective logistic and truncated distribution components.

Formally, let $y$ be a discrete non-negative response variable and let $\textbf{X}$ and $\textbf{Z}$ be two covariates matrices (that could coincide, at least partially, or be completely different), then a generic hurdle model for each individual $i$ can be defined as:
\begin{equation} \label{gen_hurdle}
    \Pr(y_i = j|\mathbf{x}_i, \mathbf{z}_i) = \left\{ \begin{array}{ll}
        f_1(0 | \mathbf{x}_i) & \mbox{for } j = 0 \\
        f_2(j|\mathbf{z}_i) \left( 1-f_1(0 |\mathbf{x}_i) \right)          & \mbox{for }j > 0      \\
        \end{array} \right.
\end{equation}
where $f_1(0|\mathbf{x}_i) = \Pr(y_i = 0|\mathbf{x}_i)$ is the probability of observing a count of 0, usually estimated from a logit or probit model, and $f_2(j|\mathbf{z}_i)$ is a truncated-at-zero count data density
\begin{equation} \label{gen_f2}
    f_2(j|\mathbf{z}_i) = \Pr(y_i = j | y_i > 0,\mathbf{z}_i) = \frac{\Pr(y_i = j |\mathbf{z}_i)}{[1 - \Pr(y_i = 0 |\mathbf{z}_i)]}
\end{equation}

The choice of the model specification for $f_2$ is usually driven by data characteristics.
In particular, one should take into account if the data are overdispersed (i.e., the variance exceeds the mean) and if there is an abnormally large number of observations concentrated on one or more values (i.e., the distribution is inflated).
As discussed in the previous section, the positive number of overnight stays is both overdispersed and inflated, therefore the specification for $f_2$ need to reflect both these characteristics.

Models that handle a single inflated value, typically at zero, have been proposed since the early 1990s, starting with the zero-inflated Poisson (ZIP) regression model first presented by \cite{Lambert1992}.
However, studies on the generalisation of single-inflated models to the situation of multiple inflated distributions are relatively recent and still in progress.

Here we move from the generalisation of the ZIP model to a multiple inflated Poisson model (MIP) suggested by \cite{Giles2007} to allow for count-inflation at multiple values.
In dealing with multiple inflated count data the MIP model assumes a finite mixture model of a Poisson distribution and a set of degenerate distributions, one for each inflated value.
In doing so, the MIP model assumes that overdispersion of data can only arise from splitting the data in more regimes.
When this assumption does not hold and the overdispersion derives also from an heterogeneity component, it is opportune to generalise the MIP model into a multiple inflated negative binomial model (MINB), in
analogy with what it is usually done when replacing a Poisson model with a negative binomial model.
Finally, since we want to model truncated-at-zero count data, the negative binomial distribution will be replaced by its truncated counterpart, obtaining a multiple inflated truncated negative binomial model (MITNB).

Assuming that the positive count response has $M-1$ inflated values, the MITNB distribution can be specified as:
   \begin{equation} \label{mitnb}
        y_i \sim \left\{ \begin{array}{ll}
        j  & \mbox{with probability $p_{ij}$ for $j=1,...,(M-1)$} \\
        TNB(\lambda_i,\theta_i) & \mbox{with probability $p_{iM}$} \\
        \end{array} \right.
    \end{equation}
where $\sum_{j=1}^{M}p_{ij}=1$ and $TNB(.)$ is the truncated negative binomial distribution.
Note that the inflated values are not required to be consecutive in the model, even if they are denoted as $1,...,(M-1)$ for notational convenience.

Under this specification, $f_2$ becomes
    \begin{equation} \label{mitnb_f2}
        \Pr(y_i=j|y_i>0,\mathbf{z}_i)=\left\{ \begin{array}{ll}
        p_{ij} + p_{iM} \dfrac{\emph{f}_{NB}(j|\mathbf{z}_i)}{[1-\emph{f}_{NB}(0|\mathbf{z}_i))]} & for j=1,...,(M-1) \\
        p_{iM} \dfrac{\emph{f}_{NB}(j|\mathbf{z}_i))}{[1-\emph{f}_{NB}(0|\mathbf{z}_i))]}          & for j \geq M      \\
        \end{array} \right.
    \end{equation}
in which
\begin{equation} \label{negbin}
            \emph{f}_{NB}(j) = \Pr(y_i = j | \mathbf{z}_i) = \dfrac{\Gamma(\lambda_i+y_i)} {\Gamma(\lambda_i)\Gamma(y_i+1)}
            \left( \dfrac{\theta_i}{1+\theta_i} \right)^{y_i} \left( \dfrac{1}{1+\theta_i}\right)^{\lambda_i}
    \end{equation}
indicates the probability mass distribution of the negative binomial model with variance function $\lambda_i(1+\lambda_i/\theta_i)$, denoted as NB2 model in \citet[p. 74]{Cameron2013}; where $\Gamma$ is the gamma function, $\lambda_i$ is the location parameter and $\theta_i$ is the scale parameter (i.e. the inverse of the dispersion parameter).

Both $\lambda_i$ and $\theta_i$ depend on covariates by the regression functions
\begin{equation}
    \ln(\lambda_i)=\mathbf{z}'_{1i} \bm{\beta}_1  \label{reg_negbin1}
\end{equation}
\begin{equation}
    \ln(\theta_i)=\mathbf{z}'_{2i}\bm{\beta}_2 \label{reg_negbin2}
\end{equation} 
where $\mathbf{Z}_1$ and $\mathbf{Z}_2$ are subsets of the covariate matrix $\mathbf{Z}$ and $\bm{\beta}_1$ and $\bm{\beta}_2$ are the vectors of the corresponding regression parameters. Note that a smaller $\theta_i$  corresponds to a larger overdispersion.

The mixing probabilities $p_{ij}=\Pr(y_i = j)$ are modelled with a multinomial logit regression model (with reference category $M$)
\begin{equation} \label{multinom}
    \ln \left[ \dfrac{\Pr(y_i = j)}{\Pr(y_i = M)} \mid \mathbf{z}_{3i} \right] = \mathbf{z}'_{3i}\bm{\gamma}_{j} 
\end{equation}
therefore
\begin{equation} \label{multinom2}
   \Pr \left( y_i = j| \mathbf{z}_{3i}\right) = 
   \dfrac{ \exp \left( \mathbf{z}'_{3i}\bm{\gamma}_{j} \right)}
         { 1 + \sum_{m=1}^{M-1} \exp \left( \mathbf{z}'_{3i}\bm{\gamma}_{m} \right) }
\end{equation}
for $j=1,...,(M-1)$, where $\mathbf{Z}_3$ is a subset of the covariate matrix $\mathbf{Z}$, $\bm{\gamma}_{j}$ is a vector of regression parameters specific to each $M-1$ value.

The multinomial logit model is an extremely flexible formulation, but requires the estimation of several parameters. If necessary one can replace it with other more parsimonious models, but these usually require additional assumptions on the parametric model formulation (for example, \cite{Su2013} use a cumulative logit model which relies on the parallel regression assumption).

Assuming, as usually done, that the error terms of the binary and the truncated model are independent, the likelihood function can be separated in two parts (one for each model component) and the two components $f_1$ and $f_2$ can be fitted separately through maximum likelihood estimation. 
Due to the model complexity, likelihood maximisation needs to be carried out by numeric optimisation techniques.

Once the MITNB hurdle model as been estimated it is possible to calculate the predicted values for the two components of the model.  
In particular, the predicted number of positive overnight stays can be computed as
\begin{equation} \label{pred_count}
\begin{split}
        & \hat E \left( y_i | y > 0, \mathbf{z}_i \right) = 
         \sum_{j=1}^{M-1} { \left( \hat p_{ij}  v_j \right) } + \hat p_{iM}  
        \hat E_{TNB} \left( y_i | y > 0, \mathbf{z}_i \right) = \\
         & = \sum_{j=1}^{M-1} { \left( \hat p_{ij}  v_j \right) } + \hat p_{iM}  
        \frac{\hat \lambda_i} {1 - \left ( 1 + \frac{\hat \lambda_i}{\hat \theta_i} \right) ^ {- \hat \theta_i} } 
\end{split}
\end{equation}
where $v_j$ is the $j$-th inflated value and $\hat p_{ij}$, $\hat \lambda_i$, $\hat \theta_i$ are respectively the mixing probabilities, the location parameter and the scale parameter predicted for a generic observation $i$ with covariates $\mathbf{z}_i$. The corresponding standard errors can then be computed by delta method.

Equation (\ref{pred_count}) highlights that, in order to accurately predict the positive values of $y$, it is essential to model not only the underline generating process (the TNB model) but also the inflated values.

\section{Results}
\label{sec:5}

This section presents the results of the empirical analysis, divided in two parts.
The first part focuses on the model specification and assessment, whereas the second part discusses the determinants of the two components (propensity and intensity) of tourism participation.

\subsection{Model specification and assessment}

The model includes a set of auxiliary variables that are collected by the \emph{Trips and Holidays}  survey for all sampled individuals (regardless of their being traveller or not). These are used as predictors of the total number of overnight stays in a quarter in the period 2004-2013.
In detail, these variables can be classified into three categories:
\begin{itemize}
    \item \textit{Socio-demographic characteristics}: gender, age (scaled; both in level and in a quadratic form), education level (academic degree vs. other levels), number of household members, presence of children up to 10 years old in the household.

    \item \textit{Economic characteristics}: the economic-related variables available in the dataset are occupational status, professional position, and number of income recipients in the household. This last variable has been transformed into relative terms and included in the model as a categorical variable: first we calculate the proportion of household members at least 16 years old whose employment status is occupied or retired, then we factorise it into three categories (``no members'', ``at most $\%50$ of members'', ``more than $\%50$ of members'').
    The individual occupational status and the professional position have been combined into a single variable that distinguishes among ``housewife/househusband'', ``student'', ``retired'', ``disabled'', ``managerial staff'', ``office worker'', ``manual worker'', ``self employed'' and ``professional''.
    We are aware that the economic condition of the individuals may not be described entirely by these variables, but unfortunately this is the only information collected by the survey.
    
    \item \textit{Temporal and spatial variables}: quarter, year (with 2004 as $t=0$; as polynomial function of order three), and regional area where the tourist lives (divided in the five Italian NUTS1 regions: ``North-west'', ``North-east'', ``Centre'', ``South'', ``Islands'').

\end{itemize}
We refer to Table \ref{tab:appendix_desc} in Appendix for the descriptive statistics of all the variables, their acronym, and definition.

Both components of the hurdle model contain the above mentioned variables as main effects. 
That is, in our analysis the covariates matrix $\mathbf{X}$ of the logit model (which governs the binary outcome  \textit{participation}/\textit{not participation in tourism}) and the covariates matrix $\mathbf{Z}$ of the truncated model (which governs the positive values of overnight stays) coincide.
The regression model for the location parameter $\lambda$ includes all the variables of matrix $\mathbf{Z}$ as well. 

About the dispersion parameter $\theta$, from Table \ref{tab_descrittive_trim} we have observed that the variability of the positive response is much higher in the third quarter, therefore to control for this heterogeneity we model $\theta$ as function of the third quarter (third vs. the others). 

Analogously, from Figure \ref{istogramma2} it is evident that the inflated values have different relevance depending on the quarter. Therefore, the mixing probabilities are estimated via a multinomial logit function of the third quarter.

The last setting required to complete the model specification is the identification of the inflated values, that is the values of overnight stays for which we observe an abnormal large frequency.
In fact, the multiple inflated model (\ref{mitnb_f2}) described in the previous section assumes that the number of inflated values ($M-1$) is known, together with their values, therefore they need to be chosen before estimating the model.
To this end, in order to identify the best model specification, we applied a two-step approach.
First, we selected a list of plausible inflated values through visual inspection of the histogram of the observed responses (Fig. \ref{istogramma}).
Then, we compared several hurdle model specifications, each characterised by a different set of inflated values, using goodness-of-fit criteria, like the Akaike information criterion (AIC) or the Bayesian information criterion (BIC) (as suggested by \cite{Cai2018}).
Table \ref{tab:setting} describes some of the considered specifications.\footnote{A larger set of model specifications has been considered in the analysis, here we present only some of them for the sake of brevity.}
In particular, the first model (Model 0) is the traditional negative binomial hurdle model without any inflated values, which has been used as benchmark model.
Models 1, 2 and 3 are all multiple inflated negative binomial hurdle models: Model 1 includes only the most evident inflated values (see Figure \ref{istogramma}), whereas Model 2 and Model 3 each adds more values to the previous model specification.
Model 4 is equivalent to Model 3, but the mixing probabilities are estimated as function of all quarters, therefore adding 32 additional parameters to the complete model.\footnote{This alternative specification has been considered for all sets of inflated values, conclusions are analogous to what presented here.}

\begin{table}[t!]
\centering
  \caption{Mixture settings in alternative specifications of the hurdle model.}
   \resizebox{\linewidth}{!}{
  \begin{tabular}{lrrrll}
   \hline\noalign{\smallskip}
    Model & $f_1$ & $f_2$ & $M-1$ & Set of inflated values         & Covariates ($\mathbf{Z}_3$) \\
    \noalign{\smallskip}\hline\noalign{\smallskip}
    Model 0 & Logit & TNB   & 0  &                                 & None         \\
    Model 1 & Logit & MITNB & 7  & (2, 6, 7, 14, 15, 20, 30)       & Quarter 3    \\
    Model 2 & Logit & MITNB & 12 & (2, 3, 4, 6, 7, 14, 15, 20, 30, & Quarter 3    \\
            &       &       &    & 40, 45, 60)                     &              \\    
    Model 3 & Logit & MITNB & 16 & (2, 3, 4, 6, 7, 10, 14, 15, 20, & Quarter 3    \\
            &       &       &    & 28, 29,  30, 40, 45, 50, 60)    &              \\    
    Model 4 & Logit & MITNB & 16 & (2, 3, 4, 6, 7, 10, 14, 15, 20, & All quarters \\
            &       &       &    & 28, 29,  30, 40, 45, 50, 60)    &              \\
   \noalign{\smallskip}\hline\noalign{\smallskip}
    \end{tabular}}\\
  \label{tab:setting}
\end{table}

\begin{table}[bt!]
\centering
  \caption{Model fit statistics for alternative specifications of the hurdle model.}
 \resizebox{.9\linewidth}{!}{
  \begin{tabular}{lrrrrrr}
   \hline\noalign{\smallskip}
    Model & $f_1$ & $f_2$ & \# parameters & logLik & AIC    & BIC  \\
    \noalign{\smallskip}\hline\noalign{\smallskip}
    Model 0 & Logit & TNB   & 60  & -333,099	&	666,318	&	666,957 \\
    Model 1 & Logit & MITNB & 74  & -329,300	&	658,748	&	659,537 \\
    Model 2 & Logit & MITNB & 84  & -328,234	&	656,635	&	657,530 \\
    Model 3 & Logit & MITNB & 92  & -327,959	&	656,101	&	657,081 \\
    Model 4 & Logit & MITNB & 124 & -327,847	&	655,942	&	657,263 \\
   \noalign{\smallskip}\hline\noalign{\smallskip}
    \end{tabular}}
  \label{tab:aic}
\end{table}

Table \ref{tab:aic} shows the corresponding model fit statistics.\footnote{Models are fitted via maximum likelihood estimation, implemented with ad hoc SAS code using PROC NLMIXED procedure (code available upon request). Maximisation is obtained via numeric optimisation applying dual quasi-Newton algorithm.} It is clear that even by considering only few extremely evident inflations, as in Model 1, we obtain a much better fit to our data comparative to the benchmark hurdle model; and the fit improves again with the addition of other inflated values.
Comparing Model 3 and Model 4 we can see that the inclusion of all quarters in the multinomial specification generates some additional gain (AIC is lower), but not as much as to be worthy of the additional complexity (BIC is higher). This confirm our choice to include only the third quarter in the multinomial model.
Therefore, the final specification for the analysis is Model 3, which considers 16 inflated values: 2, 3, 4, 6, 7, 10, 14, 15, 20, 28, 29, 30, 40, 45, 50, 60 nights.

\begin{figure} [!p]
\centering
 \includegraphics[width=.82\textwidth]{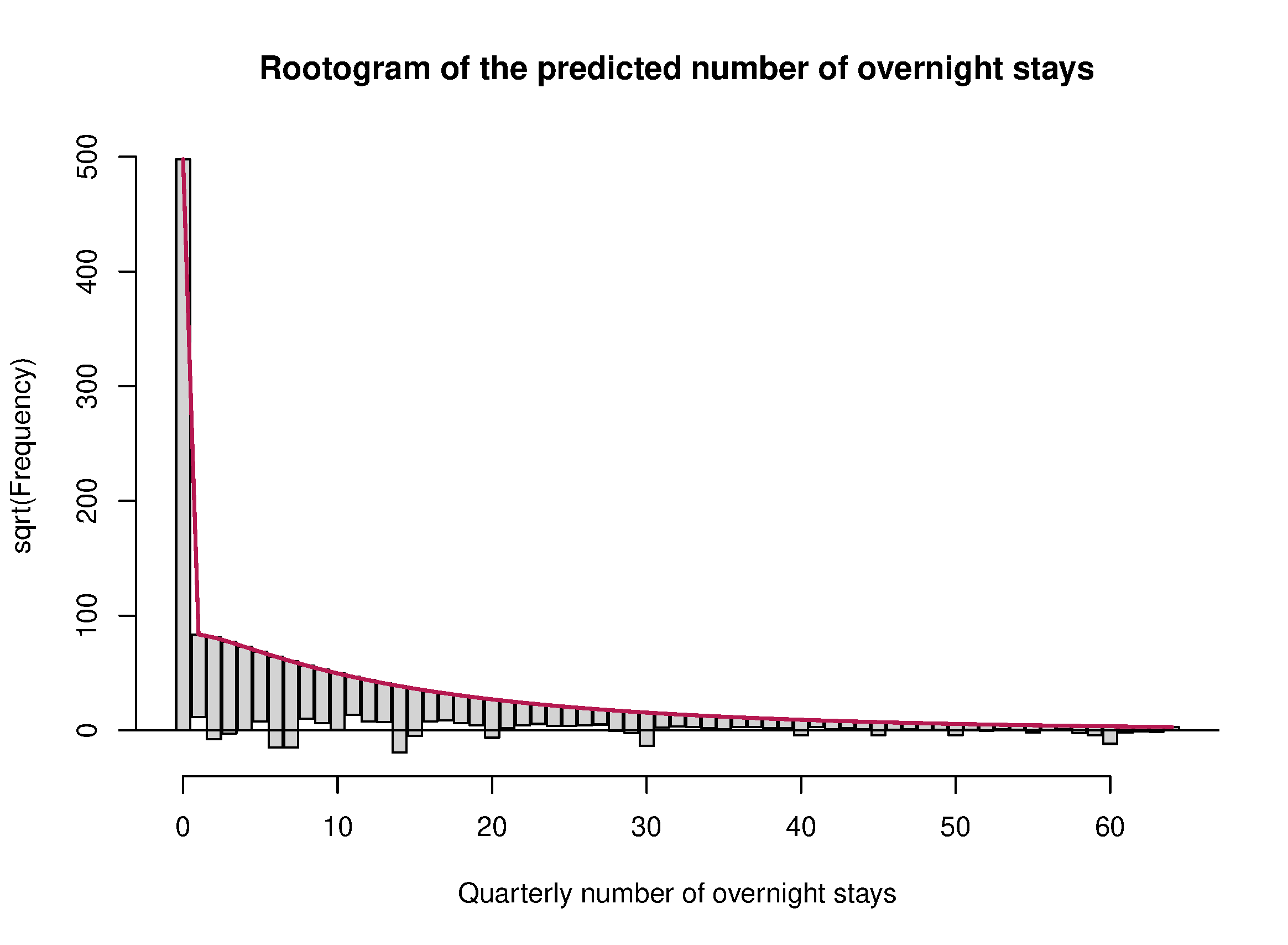}
\caption{Comparison of predicted distribution (red line) and observed distribution (hung bars), Model 0. if a bar doesn't reach the zero line indicates over-prediction, if it exceeds the zero line indicates under-prediction.}  \label{fig_rootNB}
\end{figure}
\begin{figure} [!p]
\centering
 \includegraphics[width=.82\textwidth]{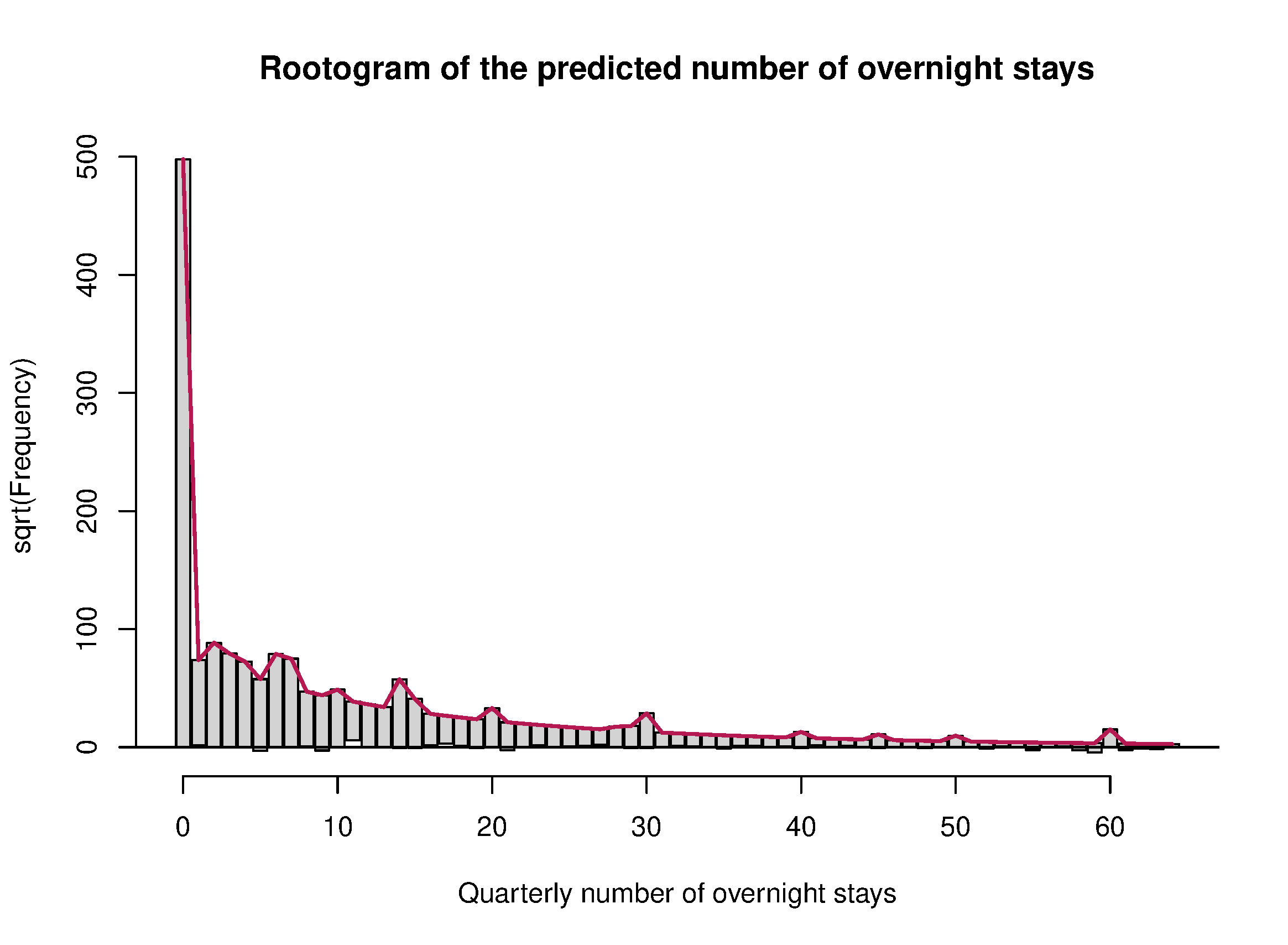}
\caption{Comparison of predicted distribution (red line) and observed distribution (hung bars), Model 3. if a bar doesn't reach the zero line indicates over-prediction, if it exceeds the zero line indicates under-prediction.}   \label{fig_root}
\end{figure}

In figures \ref{fig_rootNB} and \ref{fig_root} we present the hanging rootogram plots for Model 0 and Model 3 respectively.
As described by \cite{Kleiber2016}, the hanging rootogram is a graphic tool particularly useful for diagnosing issues such as overdispersion and multiple inflation in count data modelling.
It displays predicted and observed distribution of the variable under study, showing how the model fits the data.
Discrepancies are seen by comparison with the horizontal axis: if a bar doesn't reach the zero line then the model over-predicts a particular value, and if the bar exceeds the zero line it under-predicts it.
The vertical axis is scaled to the square-root of the frequencies to draw more attention to differences in the tails of the distribution.
The comparison between the two plots confirms that the proposed multiple inflated approach provides a better adaptation to the data since the model corrects most of the under-prediction of the inflated values that is displayed in figure \ref{fig_rootNB}.

\subsection{Determinants of tourism behaviour}

Maximum likelihood estimates of the multiple inflated hurdle model parameters are presented in the following tables: estimates for the logit regression are in Table \ref{tab_res1}, for the truncated negative binomial model are in Table \ref{tab_res2}, and for the multinomial regression are in \ref{tab_res3}.

\begin{table} [!tb]
\caption{ML estimates of logit model coefficients} \label{tab_res1} 
\center
 \resizebox{.9\linewidth}{!}{
\begin{tabular}{lr@{}lrrlr@{}l}
    \hline\noalign{\smallskip}
    \multicolumn{1}{l}{Covariate} & Coef. &&& & \multicolumn{1}{l}{Covariate} & Coef. & \\
    \noalign{\smallskip}\hline\noalign{\smallskip}
    Intercept                   &   -2.148  &   $^{***}$   &&& \emph{Prop. income recipients}   &  &              \\
                                &           &              &&& \hspace{.5cm} 0          & \multicolumn{2}{c}{ref.} \\
    Scaled age                  &   -0.529  &   $^{***}$   &&& \hspace{.5cm} (0 - 0.5]    &   0.451   &  $^{***}$    \\
    (Scaled age)$^2$            &   -0.337  &   $^{***}$   &&& \hspace{.5cm} (0.5 - 1]    &   0.677   &  $^{***}$    \\ 
    Female                      &    0.045  &   $^{***}$   &&&                          &           &              \\
    Household size              &   -0.090  &   $^{***}$   &&& \emph{NUTS1 region}     &           &              \\
    Children                    &    0.275  &   $^{***}$   &&& \hspace{.5cm} North-West & \multicolumn{2}{c}{ref.} \\
    University degree           &    0.730  &   $^{***}$   &&& \hspace{.5cm} North-East &  -0.147   &  $^{***}$    \\
    Business trips              &    0.423  &   $^{***}$   &&& \hspace{.5cm} Centre     &	 -0.230   &  $^{***}$    \\
                                &           &              &&& \hspace{.5cm} South      &  -0.682   &  $^{***}$    \\
    \emph{Occupation}          &           &              &&& \hspace{.5cm} Islands    &  -0.750   &  $^{***}$    \\
    \hspace{.5cm} Unemployed    & \multicolumn{2}{c}{ref.} &&&                          &           &              \\
    \hspace{.5cm} Housewife     &   0.174   &   $^{***}$   &&& \emph{Quarter}          &           &              \\
    \hspace{.5cm} Student      &   0.849   &   $^{***}$   &&& \hspace{.5cm} 1. Jan. - Mar.  & \multicolumn{2}{c}{ref.} \\
    \hspace{.5cm} Retired       &   0.346   &   $^{***}$   &&& \hspace{.5cm} 2. Apr. - June  &  0.366 &  $^{***}$  \\
    \hspace{.5cm} Disabled      &  -0.131   &   $^{**}$    &&& \hspace{.5cm} 3. July - Sept. &  1.381 &  $^{***}$  \\ 
    \hspace{.5cm} Managerial staff & 0.815  &   $^{***}$   &&& \hspace{.5cm} 4. Oct. - Dec.  & -0.110 &  $^{***}$  \\
    \hspace{.5cm} Office worker &   0.527   &   $^{***}$   &&&                          &           &              \\
    \hspace{.5cm} Manual worker &  -0.141   &   $^{***}$   &&& Year $t$ ($t_{2004}=0$)    &   0.175   &   $^{***}$   \\
    \hspace{.5cm} Self employed &   0.246   &   $^{***}$   &&& Year $t^2$                 &  -0.040   &   $^{***}$   \\
    \hspace{.5cm} Professional  &   0.648   &   $^{***}$   &&& Year $t^3$                 &   0.002   &   $^{***}$   \\
    \noalign{\smallskip}\hline\noalign{\smallskip}
    \end{tabular}}\\
    \emph{Significance codes}:  $^{***}$ $p<0.001$, $^{**}$ $p<0.01$, $^{*}$ $p<0.05$, $^\circ$ $p<0.1$ \\
\end{table}

\begin{table} [!tb]
\caption{ML estimates of the truncated negative binomial model coefficients} \label{tab_res2}
\center
 \resizebox{.9\linewidth}{!}{
\begin{tabular}{lr@{}lrrlr@{}l}
    \hline\noalign{\smallskip}
    \multicolumn{1}{l}{Covariate} & Coef. & &&& \multicolumn{1}{l}{Covariate} & Coef. & \\
    \noalign{\smallskip}\hline\noalign{\smallskip}
    \multicolumn{8}{l}{Model for location parameter $\vec{\lambda}$} \\
    \hline\noalign{\smallskip}
    Intercept                   &   2.221	&	 $^{***}$  &&& \emph{Prop. income recipients}   &  &              \\
                                &           &              &&& \hspace{.5cm} 0          & \multicolumn{2}{c}{ref.} \\
    Scaled age                  &  0.172	&	 $^{***}$  &&& \hspace{.5cm} (0 - 0.5]    &  -0.109  &   $^{**}$     \\
    (Scaled age)$^2$            &  0.046	&	 $^{***}$  &&& \hspace{.5cm} (0.5 - 1]    &  -0.115  &   $^{**}$     \\ 
    Female                      &  0.028	&	 $^{**}$   &&&                          &          &               \\
    Household size              & -0.074	&	 $^{***}$  &&& \emph{NUTS1 region}     &          &               \\
    Children                    &  0.217	&	 $^{***}$  &&& \hspace{.5cm} North-West & \multicolumn{2}{c}{ref.} \\
    University degree           &  0.168	&	 $^{***}$  &&& \hspace{.5cm} North-East &  -0.142	 &   $^{***}$    \\
    Business trips              & -0.022	&              &&& \hspace{.5cm} Centre     &  -0.152	 &   $^{***}$    \\
                                &           &              &&& \hspace{.5cm} South      &  -0.283  &	 $^{***}$    \\
    \emph{Occupation}          &           &              &&& \hspace{.5cm} Islands    &  -0.265  &	 $^{***}$    \\
    \hspace{.5cm} Unemployed    & \multicolumn{2}{c}{ref.} &&&                          &          &               \\
    \hspace{.5cm} Housewife     &  0.031	&              &&& \emph{Quarter}          &          &               \\
    \hspace{.5cm} Student       &  0.178	&	 $^{***}$  &&& \hspace{.5cm} 1. Jan. - Mar.  & \multicolumn{2}{c}{ref.} \\
    \hspace{.5cm} Retired       &  0.117	&	 $^{**}$   &&& \hspace{.5cm} 2. Apr. - June  & -0.035  &  $^{*}$   \\
    \hspace{.5cm} Disabled      &  0.040	&              &&& \hspace{.5cm} 3. July - Sept. &  0.828  &  $^{***}$ \\ 
    \hspace{.5cm} Managerial staff & 0.028	&              &&& \hspace{.5cm} 4. Oct. - Dec.  & -0.186	 &  $^{***}$ \\
    \hspace{.5cm} Office worker & -0.080	&	 $^{*}$    &&&                          &          &               \\
    \hspace{.5cm} Manual worker & -0.226	&	 $^{***}$  &&& Year $t$ ($t_{2004}=0$)  &  -0.051	 &	 $^{***}$    \\
    \hspace{.5cm} Self employed & -0.161	&	 $^{***}$  &&& Year $t^2$                 &   0.009	 &	 $^{*}$      \\
    \hspace{.5cm} Professional  & -0.031	&              &&& Year $t^3$                 &  -0.001	 &	 $^{**}$     \\
    \noalign{\smallskip}\hline\noalign{\smallskip}
    \multicolumn{8}{l}{Model for dispersion parameter $\vec{\theta}$} \\
    \hline\noalign{\smallskip}
    Intercept               &  0.109   & $^{***}$     &&& \emph{Quarter}             &           &              \\
                            &          &              &&& \hspace{.5cm} Quarters 1, 2, 4  & \multicolumn{2}{c}{ref.} \\
                            &          &              &&& \hspace{.5cm} 3. July - Sept. &  0.368  &  $^{***}$    \\
    \noalign{\smallskip}\hline\noalign{\smallskip}
    \end{tabular}}\\
    \emph{Significance codes}:  $^{***}$ $p<0.001$, $^{**}$ $p<0.01$, $^{*}$ $p<0.05$, $^\circ$ $p<0.1$ \\
\end{table}

\begin{table} [!tb]
\caption{ML estimates of the multinomial logit model coefficients} \label{tab_res3}
\center
 \resizebox{.75\linewidth}{!}{
\begin{tabular}{lrlr@{}lrlr@{}l}
    \hline\noalign{\smallskip}
    Inflated &&   &   && &  &   & \\
    Value && \multicolumn{1}{l}{Covariate} & Coef. && & \multicolumn{1}{l}{Covariate} & Coef. & \\
    \noalign{\smallskip}\hline\noalign{\smallskip}
    && \multicolumn{3}{l}{\emph{Intercept}:}             &&   \multicolumn{3}{l}{\emph{Quarter}:}     \\
    2	&&	$\gamma_{2,0}$	&	-2.360	&	 $^{***}$ 	&&	$\gamma_{2,Q3}$     &	-1.634	&	 $^{***}$ \\
    3	&&	$\gamma_{3,0}$	&	-2.744	&	 $^{***}$ 	&&	$\gamma_{3,Q3}$	    &	-1.765	&	 $^{***}$ \\ 
    4	&&	$\gamma_{4,0}$	&	-3.011	&	 $^{***}$ 	&&	$\gamma_{4,Q3}$	    &	-1.994	&	 $^{***}$ \\
    6	&&	$\gamma_{6,0}$	&	-2.814	&	 $^{***}$ 	&&	$\gamma_{6,Q3}$	    &	 0.372	&	 $^{***}$ \\
    7	&&	$\gamma_{7,0}$	&	-3.030	&	 $^{***}$ 	&&	$\gamma_{7,Q3}$	    &	 0.625	&	 $^{***}$ \\
    10	&&	$\gamma_{10,0}$	&	-5.083	&	 $^{***}$ 	&&	$\gamma_{10,Q3}$	&	 1.447	&	 $^{***}$ \\
    14	&&	$\gamma_{14,0}$	&	-4.120	&	 $^{***}$ 	&&	$\gamma_{14,Q3}$	&	 1.754	&	 $^{***}$ \\
    15	&&	$\gamma_{15,0}$	&	-5.535	&	 $^{***}$ 	&&	$\gamma_{15,Q3}$	&	 2.139	&	 $^{***}$ \\
    20	&&	$\gamma_{20,0}$	&	-5.472	&	 $^{***}$ 	&&	$\gamma_{20,Q3}$	&	 1.767	&	 $^{***}$ \\
    28	&&	$\gamma_{28,0}$	&	-6.744	&	 $^{***}$ 	&&	$\gamma_{28,Q3}$	&	 0.978	&	 $^{**}$  \\
    29	&&	$\gamma_{29,0}$	&	-6.894	&	 $^{***}$ 	&&	$\gamma_{29,Q3}$	&	 1.695	&	 $^{***}$ \\
    30	&&	$\gamma_{30,0}$	&	-5.736	&	 $^{***}$ 	&&	$\gamma_{30,Q3}$	&	 2.192	&	 $^{***}$ \\
    40	&&	$\gamma_{40,0}$	&	-6.723	&	 $^{***}$ 	&&	$\gamma_{40,Q3}$	&	 1.101	&	 $^{***}$ \\
    45	&&	$\gamma_{45,0}$	&	-7.442	&	 $^{***}$ 	&&	$\gamma_{45,Q3}$	&	 1.680	&	 $^{***}$ \\
    50	&&	$\gamma_{50,0}$	&	-7.886	&	 $^{***}$ 	&&	$\gamma_{50,Q3}$	&	 2.106	&	 $^{***}$ \\
    60	&&	$\gamma_{60,0}$	&	-6.781	&	 $^{***}$ 	&&	$\gamma_{60,Q3}$	&	 2.110	&	 $^{***}$ \\
    \noalign{\smallskip}\hline\noalign{\smallskip}
    \end{tabular}}\\
    \small \hspace{1cm} \emph{Significance codes}:  $^{***}$ $p<0.001$, $^{**}$ $p<0.01$, $^{*}$ $p<0.05$, $^\circ$ $p<0.1$ \\ \hspace{-4.7cm} \emph{Reference level}: Quarters 1, 2, 4
\end{table}

Results show the importance of the socio-demographic variables as determinants of both the propensity and the intensity of tourism participation. 
Age has a discordant effect: older people tend to participate less in tourism, but when they do participate they tend to have longer holidays. 
Conversely, the effect of family composition is concordant in both components of the hurdle model: a larger family has a lower propension to travel and tend to spend fewer days on holiday, but having at least one young child increases both the odds of travelling and the number of overnight stays.

Consistently with the hypothesis that economic conditions matter, the proportion of household's income recipients has a positive and highly significant effect on the decision to go on holiday. 
Moreover, estimates for the occupational status tell us that manual workers and unemployed persons are less likely to go on holiday, contrary to professionals, managerial staff and office workers who have a higher propensity to travel. 
But when on holiday, the occupational status acts primarily as a constraint on trips' duration: working individuals have less time to spend on holidays than students and retired people.
The same can be said for the proportion of income recipients, since a higher proportion is associated with fewer days of holiday. 

The model presents a remarkable North-South divide in tourism participation: assuming that the other covariates remain constant, the odds for residents of insular and southern regions are about 50\% lower than that of north-western residents. And the same North-South dualism can be observed in the number of nights spent on holiday.
In this respect, one should also consider that northern regions have a more efficient transportation system and a more favourable location as they are closer to foreign destinations that produce additional attractions for those Italian residents. 
On the other hand, since southern and insular regions have plenty of ``in-house'' leisure destinations, there could be a larger part of residents of these areas who prefer same-day trips, which are not registered in the dataset.

The estimated multinomial logit model for the mixing probabilities confirms a strong connection of the third quarter with the presence of inflated values in the distribution of the total number of overnight stays, as observed in Figure \ref{istogramma2}.
To understand the contribution of each inflated value to the mixture with the truncated negative binomial model, it is useful to calculate the mixing probabilities from the coefficients of Table \ref{tab_res3}. These probabilities, specific for years and quarters, are presented in Table \ref{tab:appendix_prob}. 

\begin{table}[!htb]
  \centering
  \caption{Estimated mixture probabilities $\Pr \left( y = j \right)$ by \emph{Quarter}. Standard errors in parenthesis, computed by delta method.}
    \center
\resizebox{.85\linewidth}{!}{
  \begin{tabular}{lrrrrrrrrrrrrrrc}
    \hline\noalign{\smallskip}
     Probability &&\multicolumn{2}{c}{Quarters 1, 2 \& 4 } && \multicolumn{2}{c}{Quarters 3 }\\
 \noalign{\smallskip}\hline\noalign{\smallskip}
    $\Pr(y = 2)$	&&	0.0697	&	(\emph{0.0024})	&&	0.0127	&	(\emph{0.0016})	\\
    $\Pr(y = 3)$	&&	0.0475	&	(\emph{0.0021})	&&	0.0076	&	(\emph{0.0015})	\\
    $\Pr(y = 4)$	&&	0.0363	&	(\emph{0.0019})	&&	0.0046	&	(\emph{0.0014})	\\
    $\Pr(y = 6)$	&&	0.0442	&	(\emph{0.0017})	&&	0.0600	&	(\emph{0.0019})	\\
    $\Pr(y = 7)$	&&	0.0357	&	(\emph{0.0015})	&&	0.0623	&	(\emph{0.0019})	\\
    $\Pr(y = 10)$	&&	0.0046	&	(\emph{0.0009})	&&	0.0182	&	(\emph{0.0013})	\\
    $\Pr(y = 14)$	&&	0.0120	&	(\emph{0.0008})	&&	0.0648	&	(\emph{0.0017})	\\
    $\Pr(y = 15)$	&&	0.0029	&	(\emph{0.0006})	&&	0.0231	&	(\emph{0.0012})	\\
    $\Pr(y = 20)$	&&	0.0031	&	(\emph{0.0004})	&&	0.0170	&	(\emph{0.0010})	\\
    $\Pr(y = 28)$	&&	0.0009	&	(\emph{0.0002})	&&	0.0022	&	(\emph{0.0005})	\\
    $\Pr(y = 29)$	&&	0.0007	&	(\emph{0.0002})	&&	0.0038	&	(\emph{0.0006})	\\
    $\Pr(y = 30)$	&&	0.0024	&	(\emph{0.0003})	&&	0.0199	&	(\emph{0.0009})	\\
    $\Pr(y = 40)$	&&	0.0009	&	(\emph{0.0002})	&&	0.0025	&	(\emph{0.0004})	\\
    $\Pr(y = 45)$	&&	0.0004	&	(\emph{0.0001})	&&	0.0022	&	(\emph{0.0004})	\\
    $\Pr(y = 50)$	&&	0.0003	&	(\emph{0.0001})	&&	0.0021	&	(\emph{0.0003})	\\
    $\Pr(y = 60)$	&&	0.0008	&	(\emph{0.0002})	&&	0.0065	&	(\emph{0.0005})	\\
    $1-\sum_{j=1}^{M-1}\Pr(y = j)$	&&	0.7376	&	(\emph{0.0058})	&&	0.6903	&	(\emph{0.0055})	\\
\noalign{\smallskip}\hline\noalign{\smallskip}
\end{tabular}}
  \label{tab:appendix_prob}
\end{table}

Overall, the mixing weights show that inflations are a non-negligible component of the intensity of tourism participation, especially in the third quarter in which less than 70\% of the positive values seems to derive from the truncated negative binomial distribution. Even during the other quarters, the percentage of positive values explained by the TNB is about 73\%. What is interesting to note, however, is the different composition of the mixture in the third quarter than in the rest of the year: the smaller inflated values have larger weights in the off-peak seasons, whereas in the third quarter the most frequent values are 6, 7, 14, 15 and 30 days. These values derive from the summation of one or more trips, which are commonly taken in weekend-, week-, or month-long blocks.

To understand the effect of the economic crisis, we compute the predictive margins of year and quarter for the two aspects of tourism participation, that is: i) the predicted probability of having at least one trip and ii) the expected number of positive overnight stays, as functions of year and quarter. Predictive margins are computed as the average of the predicted values for all observations at each fixed value of year and quarter (leaving other covariates at their observed value). Since year and quarter influence each part of the MITNB hurdle model, by calculating the predictive margin we are able to see the overall role of the two covariates.

\begin{figure} [!p]
\centering
\hspace{-1cm}\includegraphics[width=.87\textwidth]{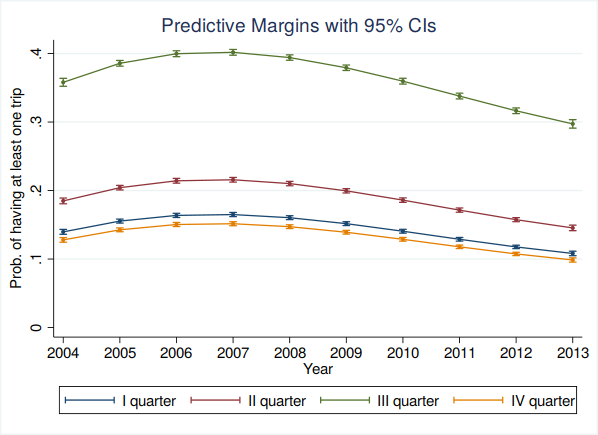}
\caption{Predictive margins of \emph{Year} and \emph{Quarter} on tourism participation} \label{fig_zero}
\end{figure}

\begin{figure} [!p]
\hspace{.4cm}\includegraphics[width=.9\textwidth]{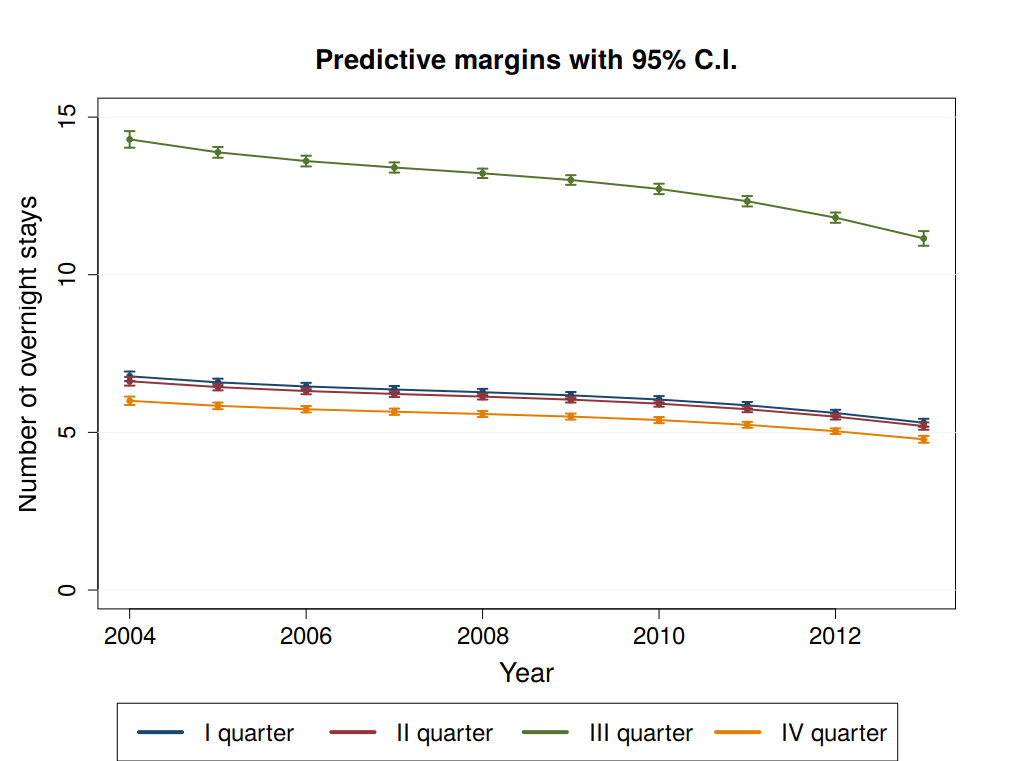}
\caption{Predictive margins of \emph{Year}, \emph{Quarter} and \emph{Short trip} on positive values of the quarterly number of overnight stays.} \label{fig_count}
\end{figure}

Predictive margins are plotted in Figures \ref{fig_zero} and \ref{fig_count} and reported in Table \ref{tab:appendix_margins} in Appendix. 
Graphs show that the reaction to the Great Recession starts in 2009 when, after a period of growth in participation (comparatively to 2004), the predicted probability of travelling begins to decrease; and the decline spikes in 2011 probably due to the heavy fiscal restrictive measures adopted by the Italian government in that year. 
In addition, from 2011 the tourism participation drops below the 2004's level and, after a year of stability, further decrease in 2013.
This might indicate the presence of an inertia in reacting to the 2008 crisis: at first, households reacted to an increase in taxes by reducing savings to defend their living standards; after, considering the persistence of the crisis, households had to reduce consumption. 
Moreover, results reflect the general trend of a reduction in the average length of stay per trip which has been observed at the macro level \cite[Chp. 18]{ISTAT2014}. In fact the decrease seams to have started even before 2008, but the highest reduction can be identified in 2013 indicating that the 2011 downturn, more than that of 2008, strongly affected tourism behaviour about intensity.

\section{Final remarks}

Estimation results for the proposed multiple inflated hurdle regression model show that, in Italy, the Great Recession had a negative impact on both the propensity and the intensity of tourism participation.
Moreover, estimates confirm common knowledge that seasonality is a universal factor in tourism and that socio-demographic and economic characteristics are relevant in determining individuals' tourism behaviour.

In assessing the effects of the Great Recession on tourism, we have to consider that nowadays tourism has become a ``normal thing'', a part of the lifestyle, quality of life and well being of an increasing number of people \citep{Bargeman2006, Dolnicar2012, Cracolici2013}.
Therefore, we should likely observe an inertia in tourism behaviour and a higher probability of a ``slicing strategy'' (e.g., cheaper holiday) rather than a pure ``cutback strategy'' (e.g., fewer trips, reduced length of stay) \citep{Bronner2012}.
The dataset used in our analysis doesn't include information about tourism expenditures, therefore we were not able to investigate whether Italians employs a ``slicing strategy'' in response to the economic crisis. Conversely, through the formulation of a hurdle model we studied which level of ``cutback strategy'' has been mostly implemented by the Italian citizens: (a) giving up holidays completely or (b) reducing in the number of overnight stays. 
Evidence shows that both strategies has been applied in the period of the Great Recession, but the more prominent reaction to the crisis seems to be the complete renounce to leisure trips: the probability of participation diminished between 2007 and 2013 by more that 30\% in the off-peak seasons and by 25\% in the summer.

Motivated by the analysis of the impact of the Great Recession on tourism behaviour, the paper proposes a general, novel approach for dealing with count variables whose distribution is inflated in multiple values. 
This feature can not be represented through the probability distribution models commonly used for count data, but needs to be properly addressed (alongside other data characteristics like zero-inflation and overdispersion) in order to avoid possible estimation biases and incorrect inference about the model parameters \citep{Cai2018}.
Moreover, failing to control for the inflated nature of the distribution can limit the model's ability to produce reliable model based predictions.

We propose the use of a multiple inflated hurdle negative binomial model, with mixing probabilities modelled through a multinomial logit model, in comparison with the use of the well known hurdle negative binomial model. 
We show that, by controlling for the inflated nature of the response variable distribution, the proposed formulation provides a better representation of the Italians' tourism behaviour in comparison with non-inflated hurdle models. 
In particular, by using a multiple inflated hurdle model we are able not only to identify the determinants of the phenomenon under study, but also to correctly fit the distribution of the total number of overnight stays, even in presence of extremely inflated values which are usually under-predicted by standard models.
Given this characteristic, we believe that multiple inflated hurdle models can be useful tools for decision makers who are trying to forecast future events or the consequences of some new targeted policies.

The proposed methodology assumes that the inflated values are known, or are exogenously selected by a double procedure of visual inspection and model comparison. 
Optimal selection of the mixture components is a controversial issue when using any mixture model, and further research should be devoted to investigate the possibility of including the identification of the inflated values directly in the model estimation process.

\newpage
\section{Appendix}

\begin{table}[!htbp]
  \centering
  \caption{Descriptive statistics}
  \resizebox{\linewidth}{!}{
  \begin{tabular}{lrrrrl}
    \hline\noalign{\smallskip}
    \multicolumn{1}{l}{Variable} & \multicolumn{1}{l}{Mean} & \multicolumn{1}{l}{Std. Deviation} & \multicolumn{1}{l}{Min} & \multicolumn{1}{l}{Max} & \multicolumn{1}{l}{Definition} \\
    \noalign{\smallskip}\hline\noalign{\smallskip}
    Age             & 52.647 & 19.831 & 15 & 109 & Age \\
    Female          &  0.533 &  0.499 &  0 &   1 & Female gender \\
    Household size  &  2.980 &  1.230 &  1 &  10 & Number of household's members \\
    Children        &  0.131 &  0.338 &  0 &   1 & At least one child up to 10 years old in \\
    & & & & & the household \\
    University Degree &  0.105 &  0.306 &  0 &   1 & University degree vs other education levels \\
    Business trips  &  0.026 &  0.160 &  0 &   1 & The individual has had business trips in \\
    & & & & & the reference quarter \\
    \emph{Occupation}  & & & & & \emph{Occupational status of the individual}\\
    \hspace{.2cm} Unemployed  &  0.038 &  0.192 &  0 &   1 & \hspace{.2cm} In search of employment \\
    \hspace{.2cm} Housewife   &  0.141 &  0.348 &  0 &   1 & \hspace{.2cm} Housewife/househusband \\
    \hspace{.2cm} Student     &  0.094 &  0.292 &  0 &   1 & \hspace{.2cm} Student \\
    \hspace{.2cm} Retired     &  0.292 &  0.455 &  0 &   1 & \hspace{.2cm} Retired \\
    \hspace{.2cm} Disabled    &  0.055 &  0.228 &  0 &   1 & \hspace{.2cm} Disabled for work \\
    \hspace{.2cm} Managerial staff &  0.026 &  0.160 &  0 & 1 & \hspace{.2cm} Employed in a management position \\
    \hspace{.2cm} Office worker &  0.164 &  0.370 &  0 &   1 & \hspace{.2cm} Employed as office worker \\
    \hspace{.2cm} Manual worker &  0.114 &  0.318 &  0 &   1 & \hspace{.2cm} Employed as manual worker \\
    \hspace{.2cm} Self employed &  0.051 &  0.221 &  0 &   1 & \hspace{.2cm} Self employed (entrepreneur or craftsman) \\
    \hspace{.2cm} Professional &  0.023 &  0.151 &  0 &   1 & \hspace{.2cm} Professional \\
    \emph{Prop. income recipients}  & & & & & \emph{Proportion of household's members employed}\\
    & & & & & \emph{or retired, at least 16 years old}\\
    \hspace{.2cm} [0]       &  0.059 &  0.237 &  0 &   1 & \hspace{.2cm} $= 0$ \\
    \hspace{.2cm} (0 - 0.5] &  0.348 &  0.476 &  0 &   1 & \hspace{.2cm} $> 0$ and $\leq 0.5$ \\
    \hspace{.2cm} (0.5 - 1] &  0.592 &  0.491 &  0 &   1 & \hspace{.2cm} $> 0.5$ \\
    \emph{NUTS1 region}  & & & & & \emph{Regional area where the tourist lives} \\ 
    \hspace{.2cm} North-West &  0.224 &  0.417 &  0 &   1 & \hspace{.2cm} North-western Italy \\
    \hspace{.2cm} North-East &  0.206 &  0.405 &  0 &   1 & \hspace{.2cm} North-eastern Italy \\
    \hspace{.2cm} Centre    &  0.194 &  0.395 &  0 &   1 & \hspace{.2cm} Central Italy \\
    \hspace{.2cm} South     &  0.269 &  0.443 &  0 &   1 & \hspace{.2cm} South Italy \\
    \hspace{.2cm} Islands   &  0.107 &  0.309 &  0 &   1 & \hspace{.2cm} Insular Italy \\
    \emph{Quarter}        & & & & & \emph{Quarter} \\
    \hspace{.2cm} 1. Jan. - Mar.  &  0.251 &  0.434 &  0 &   1 & \hspace{.2cm} First quarter: January - March \\
    \hspace{.2cm} 2. Apr. - June  &  0.252 &  0.434 &  0 &   1 & \hspace{.2cm} Second quarter: April - June   \\
    \hspace{.2cm} 3. July - Sept. &  0.249 &  0.432 &  0 &   1 & \hspace{.2cm} Third quarter: July - September \\
    \hspace{.2cm} 4. Oct. - Dec.  &  0.248 &  0.432 &  0 &   1 & \hspace{.2cm} Fourth quarter: October - December \\
    Year $t$        &  4.386 &  2.859 &  0 &   9 & Years from 2004 ($t_{2004} = 0$) \\
    \noalign{\smallskip}\hline\noalign{\smallskip}
        \end{tabular}}
  \label{tab:appendix_desc}
\end{table}

\begin{table}[!htbp]
  \centering
  \caption{Predictive margins of \emph{Year} and \emph{Quarter} on the two components of tourism behaviour: probability $\hat p(y>0)$ and intensity $\hat E(y|y>0)$ of tourism participation. Standard errors in parenthesis, computed by delta method.}
   \resizebox{.48\linewidth}{!}{
  \begin{tabular}{llrllrllrl}
    \hline\noalign{\smallskip}
\multicolumn{1}{c}{Year}	& &	\multicolumn{2}{c}{$\hat p(y>0)$} & &	\multicolumn{2}{c}{$\hat E(y|y>0)$}	\\
     \hline\noalign{\smallskip}
\multicolumn{7}{c}{Quarter 1} \\
\hline\noalign{\smallskip}
2004	&&	0.140	&	(\emph{0.002})	&&	6.778	&	(\emph{0.078})	\\
2005	&&	0.155	&	(\emph{0.001})	&&	6.586	&	(\emph{0.060})	\\
2006	&&	0.164	&	(\emph{0.002})	&&	6.456	&	(\emph{0.059})	\\
2007	&&	0.165	&	(\emph{0.002})	&&	6.360	&	(\emph{0.057})	\\
2008	&&	0.160	&	(\emph{0.001})	&&	6.274	&	(\emph{0.054})	\\
2009	&&	0.152	&	(\emph{0.001})	&&	6.175	&	(\emph{0.054})	\\
2010	&&	0.141	&	(\emph{0.001})	&&	6.042	&	(\emph{0.055})	\\
2011	&&	0.129	&	(\emph{0.001})	&&	5.859	&	(\emph{0.054})	\\
2012	&&	0.118	&	(\emph{0.001})	&&	5.615	&	(\emph{0.052})	\\
2013	&&	0.108	&	(\emph{0.002})	&&	5.306	&	(\emph{0.064})	\\
\noalign{\smallskip}\hline\noalign{\smallskip}
\multicolumn{7}{c}{Quarter 2} \\
\hline\noalign{\smallskip}
2004	&&	0.185	&	(\emph{0.002})	&&	6.621	&	(\emph{0.070})	\\
2005	&&	0.204	&	(\emph{0.002})	&&	6.435	&	(\emph{0.053})	\\
2006	&&	0.214	&	(\emph{0.002})	&&	6.310	&	(\emph{0.052})	\\
2007	&&	0.216	&	(\emph{0.002})	&&	6.218	&	(\emph{0.050})	\\
2008	&&	0.210	&	(\emph{0.002})	&&	6.135	&	(\emph{0.048})	\\
2009	&&	0.200	&	(\emph{0.002})	&&	6.039	&	(\emph{0.048})	\\
2010	&&	0.186	&	(\emph{0.002})	&&	5.910	&	(\emph{0.049})	\\
2011	&&	0.171	&	(\emph{0.002})	&&	5.734	&	(\emph{0.048})	\\
2012	&&	0.157	&	(\emph{0.002})	&&	5.498	&	(\emph{0.047})	\\
2013	&&	0.145	&	(\emph{0.002})	&&	5.200	&	(\emph{0.059})	\\
\noalign{\smallskip}\hline\noalign{\smallskip}
\multicolumn{7}{c}{Quarter 3} \\
\hline\noalign{\smallskip}
2004	&&	0.358	&	(\emph{0.003})	&&	14.293	&	(\emph{0.134})	\\
2005	&&	0.386	&	(\emph{0.002})	&&	13.882	&	(\emph{0.087})	\\
2006	&&	0.400	&	(\emph{0.002})	&&	13.605	&	(\emph{0.086})	\\
2007	&&	0.402	&	(\emph{0.002})	&&	13.401	&	(\emph{0.082})	\\
2008	&&	0.394	&	(\emph{0.002})	&&	13.217	&	(\emph{0.076})	\\
2009	&&	0.379	&	(\emph{0.002})	&&	13.005	&	(\emph{0.078})	\\
2010	&&	0.360	&	(\emph{0.002})	&&	12.721	&	(\emph{0.084})	\\
2011	&&	0.338	&	(\emph{0.002})	&&	12.331	&	(\emph{0.084})	\\
2012	&&	0.316	&	(\emph{0.002})	&&	11.810	&	(\emph{0.083})	\\
2013	&&	0.297	&	(\emph{0.003})	&&	11.152	&	(\emph{0.118})	\\
\noalign{\smallskip}\hline\noalign{\smallskip}
\multicolumn{7}{c}{Quarter 4} \\
\hline\noalign{\smallskip}
2004	&&	0.128	&	(\emph{0.002})	&&	6.001	&	(\emph{0.068})	\\
2005	&&	0.143	&	(\emph{0.001})	&&	5.841	&	(\emph{0.053})	\\
2006	&&	0.150	&	(\emph{0.002})	&&	5.734	&	(\emph{0.052})	\\
2007	&&	0.152	&	(\emph{0.001})	&&	5.655	&	(\emph{0.051})	\\
2008	&&	0.147	&	(\emph{0.001})	&&	5.583	&	(\emph{0.049})	\\
2009	&&	0.139	&	(\emph{0.001})	&&	5.501	&	(\emph{0.049})	\\
2010	&&	0.129	&	(\emph{0.001})	&&	5.391	&	(\emph{0.049})	\\
2011	&&	0.118	&	(\emph{0.001})	&&	5.239	&	(\emph{0.048})	\\
2012	&&	0.107	&	(\emph{0.001})	&&	5.036	&	(\emph{0.047})	\\
2013	&&	0.099	&	(\emph{0.002})	&&	4.780	&	(\emph{0.056})	\\
\noalign{\smallskip}\hline\noalign{\smallskip}
\end{tabular}}
  \label{tab:appendix_margins}
\end{table}

\begin{table} [!htbp]           
\caption{ML estimates of the negative binomial model, Model 0.} \label{tab_mod0}    
\center     
\resizebox{.9\linewidth}{!}{
\begin{tabular}{lr@{}lrrlr@{}l}
    \hline\noalign{\smallskip}
    \multicolumn{1}{l}{Covariate} & Coef. & &&& \multicolumn{1}{l}{Covariate} & Coef. & \\
    \noalign{\smallskip}\hline\noalign{\smallskip}
    \multicolumn{8}{l}{Model for location parameter $\vec{\lambda}$} \\
    \hline\noalign{\smallskip}
    Intercept & 2.147 & $^{***}$ &&& \emph{Prop. income recipients} & & \\
              &       &          &&& \hspace{.5cm} 0     & \multicolumn{2}{c}{ref.} \\
    Scaled age & 0.162 & $^{***}$ &&& \hspace{.5cm} (0 - 0.5]   & -0.084 & $^{**}$  \\
    (Scaled age)$^2$  & 0.047 &   $^{***}$  &&&  \hspace{.5cm} (0.5 - 1]     & -0.090 &   $^{**}$    \\
    Female            & 0.024 &   $^{**}$   &&&                      &            &  \\
    Household size    & -0.057 &   $^{***}$ &&&  \emph{NUTS1 region}       &            &  \\
    Children          & 0.182 &   $^{***}$  &&&  \hspace{.5cm} North-West  &  \multicolumn{2}{c}{ref.} \\
    University degree & 0.135 &   $^{***}$  &&&  \hspace{.5cm} North-East  & -0.123 &   $^{***}$   \\
    Business trips    & -0.033 & $^\circ$   &&&  \hspace{.5cm} Centre      & -0.121 &   $^{***}$   \\
                      &        &            &&&  \hspace{.5cm} South       & -0.201 &   $^{***}$   \\
    \emph{Occupation} &        &            &&&  \hspace{.5cm} Islands     & -0.170 &   $^{***}$   \\
    \hspace{.5cm} Unemployed &  \multicolumn{2}{c}{ref.} &&&            &  &  \\
    \hspace{.5cm} Housewife  & -0.006 &     &&&  \emph{Quarter}           &            &  \\
    \hspace{.5cm} Student &  0.085 & $^{***}$ &&& \hspace{.5cm} 1. Jan. - Mar. & \multicolumn{2}{c}{ref.} \\
    \hspace{.5cm} Retired &  0.060 & $^{*}$  &&&  \hspace{.5cm} 2. Apr. - June   & -0.041 &   $^{**}$  \\
    \hspace{.5cm} Disabled & 0.013 &       &&&  \hspace{.5cm} 3. July - Sept.  & 0.844 &   $^{***}$   \\
    \hspace{.5cm} Managerial staff  & -0.045 &  &&&  \hspace{.5cm} 4. Oct. - Dec. & -0.167 & $^{***}$   \\
    \hspace{.5cm} Office worker  & -0.123 &   $^{***}$   &&&           &  &  \\
    \hspace{.5cm} Manual worker  & -0.216 &   $^{***}$ &&&  Year $t$ ($t_{2004}=0$) & -0.039 & $^{***}$ \\
    \hspace{.5cm} Self employed  & -0.174 &   $^{***}$ &&&  Year $t^2$   & 0.007 &   $^{*}$     \\
    \hspace{.5cm} Professional   & -0.088 &   $^{**}$  &&&  Year $t^3$   & -0.001 &   $^{**}$    \\
    \noalign{\smallskip}\hline\noalign{\smallskip}
    \multicolumn{8}{l}{Model for dispersion parameter $\vec{\theta}$} \\
    \hline\noalign{\smallskip}
    Intercept   &  0.203   & $^{***}$     &&& \emph{Quarter}             &           &              \\
                &          &              &&& \hspace{.5cm} Quarters 1, 2, 4  & \multicolumn{2}{c}{ref.} \\
                &          &              &&& \hspace{.5cm} 3. July - Sept. &  0.373  &  $^{***}$    \\
    \noalign{\smallskip}\hline\noalign{\smallskip}           
    \end{tabular}}\\           
    \emph{Significance codes}:  $^{***}$ $p<0.001$, $^{**}$ $p<0.01$, $^{*}$ $p<0.05$, $^\circ$ $p<0.1$ \\           
\end{table}           

\newpage
\bibliography{paper_arXiv}
\bibliographystyle{agsm}

\end{document}